\documentclass[aps,prx,twocolumn,footinbib,showpacs,showkeys]{revtex4-1}
\usepackage{graphicx}
\usepackage{indentfirst}
\usepackage{physics}
\usepackage{braket}
\usepackage{float}
\usepackage{amsmath}
\usepackage{epstopdf}
\usepackage{CJK}
\usepackage{esint}
\usepackage{color}
\usepackage[T1]{fontenc}
\usepackage{subfigure}
\usepackage{amsfonts}
\usepackage{footmisc}
\usepackage{scrextend}
\usepackage{multirow}
\usepackage[hyperfootnotes=false]{hyperref}

\usepackage[english]{babel}
\usepackage{url}
\usepackage{bm}
\usepackage{hyperref}
\definecolor{darkblue}{rgb}{0,0,0.5}
\hypersetup{
colorlinks=true,
linkcolor=black,
filecolor=blue,
citecolor=darkblue,  
urlcolor=black,
}

\newcommand\argmin{\mathop{\mathrm{argmin}}}
\newcommand{\calA}{{\cal A}}

\newcommand{\calE}{{\cal E}}

\newcommand{\calL}{{\cal L}}
\newcommand{\calN}{{\cal N}}

\newcommand{\1}{^{(1)}}

\def\be{\begin{equation}}
\def\ee{\end{equation}}
\def\ba{\begin{eqnarray}}
\def\ea{\end{eqnarray}}
\usepackage{bm}

\begin{document}

\title{Distributed quantum sensing enhanced by continuous-variable error correction}

\author{Quntao Zhuang$^{1,2}$}
\email{zhuangquntao@email.arizona.edu}
\author{John Preskill$^{3,4}$}
\author{Liang Jiang$^{5,6}$}
\affiliation{
$^1$Department of Electrical and Computer Engineering, University of Arizona, Tucson, Arizona 85721, USA
\\
$^2$James C. Wyant College of Optical Sciences, University of Arizona, Tucson, AZ 85721, USA
\\
$^3$
Institute for Quantum Information and Matter, Caltech, Pasadena, CA, USA
\\
$^4$
Walter Burke Institute for Theoretical Physics, Caltech, Pasadena, CA, USA
\\
$^5$Pritzker School of Molecular Engineering, University of Chicago, 5640 S. Ellis Ave., Chicago, Illinois 60637, United States
\\
$^6$Yale Quantum Institute, Departments of Applied Physics and Physics, Yale University, New Haven, CT 06520, United States of America
}
\date{\today}

\begin{abstract}
A distributed sensing protocol uses a network of local sensing nodes to estimate a global feature of the network, such as a weighted average of locally detectable parameters. 
In the noiseless case, continuous-variable multipartite entanglement shared by the nodes can improve the precision of parameter estimation relative to the precision attainable by a network without shared entanglement; for an entangled protocol, the root-mean-square estimation error scales like $1/M$ with the number $M$ of sensing nodes, the so-called Heisenberg scaling, while for protocols without entanglement, the error scales like $1/\sqrt{M}$. However, in the presence of loss and other noise sources, although multipartite entanglement still has some advantages for sensing displacements and phases, the scaling of the precision with $M$ is less favorable. In this paper, we show that using continuous-variable error correction codes can enhance the robustness of sensing protocols against imperfections and reinstate Heisenberg scaling up to moderate values of $M$. 
Furthermore, while previous distributed sensing protocols could measure only a single quadrature, we construct a protocol in which both quadratures can be sensed simultaneously. 
Our work demonstrates the value of continuous-variable error correction codes in realistic sensing scenarios.
\end{abstract} 


\maketitle

Quantum sensing~\cite{Helstrom_1976,caves1981quantum,Braunstein1994,Giovannetti_2001,giovannetti2004,giovannetti2006,paris2009quantum,giovannetti2011advances,pirandola2018advances,Braun2018,lawrie2019quantum} uses nonclassical resources to enhance measurement precision. It has many applications, including atomic clocks~\cite{mcgrew2018atomic,newman2019architecture}, the Laser Interferometer Gravitational-Wave Observatory (LIGO)~\cite{LIGO_nat,LIGO}, quantum illumination~\cite{Lloyd2008,Tan2008,Guha2009,Zhang2013,barzanjeh2015microwave,Zheshen_15,Zhuang2017}, quantum reading~\cite{pirandola2011quantum} and bio-sensing~\cite{taylor2013biological}. When the sensing task involves multiple parties, entanglement can be extremely beneficial.
Early works have already shown that when measuring a single physical parameter with $M$ sensor probes, entanglement among the sensors can reduce the root-mean-square (rms) estimation error to the Heisenberg scaling~\cite{Giovannetti_2001,giovannetti2004,giovannetti2006,zwierz2010general,zhang2014fitting,komar2014quantum,Humphreys_2013} of $\propto1/M$. In contrast, in the absence of entanglement, the rms estimation error always obeys the standard quantum limit (SQL) scaling of $\propto1/\sqrt{M}$, as dictated by the law of large numbers.

More recently, this separation between Heisenberg and SQL scaling has been generalized to the scenario of distributed sensing, where an array of sensors aims to sense a global feature, such as a weighted average, of some local parameters detected by different sensor nodes~\cite{proctor2017multi,ge2017distributed,zhuang2018distributed,eldredge2018optimal,qian2019heisenberg}.
In particular, Ref.~\cite{zhuang2018distributed} proposed a protocol to use continuous variable (CV) multi-partite entanglement to enhance the distributed sensing of displacements and phases, which led to the first experimental demonstration~\cite{guo2019distributed} of sensing advantage enabled by multi-partite entanglement.

Despite being more robust against loss than their discrete variable (DV) cousins, the performance enhancement in CV distributed sensing protocols still decays in the presence of loss and noise~\cite{escher2011}. As a consequence, Ref.~\cite{guo2019distributed} only achieved a $\sim20\%$ advantage in the rms estimation error. Therefore, loss mitigation is crucial for achieving a practical advantage in distributed sensing. To this end, Ref.~\cite{xia2018repeater} proposed to use nonlinear amplifiers~\cite{ralph2009nondeterministic} to non-deterministically reduce loss in the state distribution process. In this paper, we propose to use the recently developed CV error correction codes~\cite{noh2019encoding} to mitigate loss and noise in a deterministic manner.

There has been various proposals to improve the sensing precision with error correction codes in DV systems~\cite{chaves2013noisy,dur2014improved,kessler2014quantum,arrad2014increasing,zhou2018achieving,layden2019ancilla,theodoros2019}. Most of these works consider a Hamiltonian parameter estimation scenario, where frequent error correction steps are applied to suppress the noise without at the same time suppressing the signal. When the ``Hamiltonian not in the Lindblad span'' (HNLS) criterion is satisfied, the Heisenberg scaling in precision can be reinstated~\cite{zhou2018achieving}. For applications like radio-frequency (RF) sensing or bio-sensing, however, the sensing process on each spatially distributed sensing node is modeled as a quantum channel. In such a distributed-channel parameter estimation scenario, the distribution loss is a major source of imperfection; we propose to use CV error correction to mitigate this loss.
Using the recently developed CV error correction codes based on the Gottesman-Kitaev-Preskill (GKP) code~\cite{gottesman2001},
we show that the measurement precision can be substantially improved when the loss is not too high. Inspired by the idea of measuring commuting operators associated with the grid state~\cite{duivenvoorden2017}, we also extend the distributed sensing protocol to simultaneously achieve the Heisenberg scaling on both quadratures. We do so without using any ancilla in the source, unlike in the usual super-dense sensing scheme~\cite{Genoni_2013}.

Our paper is organized as follows. In section~\ref{Sec_distributed_sensing}, we introduce the distributed sensing protocol for real quadrature displacements. In section~\ref{Sec_EC} we introduce the CV error correction codes, including the GKP-two-mode-squeezing code in section~\ref{Sec_EC_tms} and the GKP-stabilizer code in section~\ref{Sec_EC_stabilizer}. Finally, in section~\ref{Sec_improvement}, we evaluate the performance improvement achieved using error correction in distributed sensing schemes, including loss mitigation in the sensing of single-quadrature displacements in section~\ref{Sec_improvement_real} and the extension to the sensing of {\em complex-valued displacement} (i.e. displacements on both quadratures) in section~\ref{Sec_improvement_complex}. 

\section{Distributed sensing of real quadrature displacements}
\label{Sec_distributed_sensing}

For CV sensors, the signal is acquired by measuring the displacement changes in the sensor state---e.g., position and/or momentum change for a mechanical oscillator. The precise measurement of displacements is important for interferometric phase sensing~\cite{escher2011}, quantum key distribution~\cite{grosshans2002continuous}, spin sensing~\cite{eckert2008quantum}, and inertia sensing~\cite{krause2012high}. Moreover, like the example in RF sensing~\cite{xia2019entangled}, transducers can transform a even broader class of signals into optical displacements for further sensing purposes. Mathematically, displacements are described by the unitary $\hat{U}(\alpha)=\exp\left(\hat{a}^\dagger\alpha-\hat{a}\alpha^\star\right)$, or equivalently the mode transform $\hat{a}\to \hat{a}+\alpha$. Here $\hat{a}$ is the annihilation operator of the field being sensed. Equivalently, a displacement $\hat{U}(\alpha)$ can also be represented by a quadrature transform $(\hat{q},\hat{p}) \to (\hat{q}+\sqrt{2}{\rm Re}(\alpha),\hat{p}+\sqrt{2}{\rm Im}(\alpha))$, where $\hat{q}=(\hat{a}^\dagger+\hat{a})/\sqrt{2}$, $\hat{p}=i(\hat{a}^\dagger-\hat{a})/\sqrt{2}$ are the position and momentum quadratures. For simplicity, we will use the notation $\bm \alpha=({\rm Re}(\alpha),{\rm Im}(\alpha))$. In this convention, the quadrature variance is $\hat{p}^2+\hat{q}^2=2\hat{n}+1$, where $\hat{n}\equiv\hat{a}^\dagger \hat{a}$ is the number operator. Thus the vacuum noise ($\langle \hat n\rangle = 0$) is $\braket{\hat{p}^2}=\braket{\hat{q}^2}=1/2$.  

\begin{figure}
\centering
\includegraphics[width=0.4\textwidth]{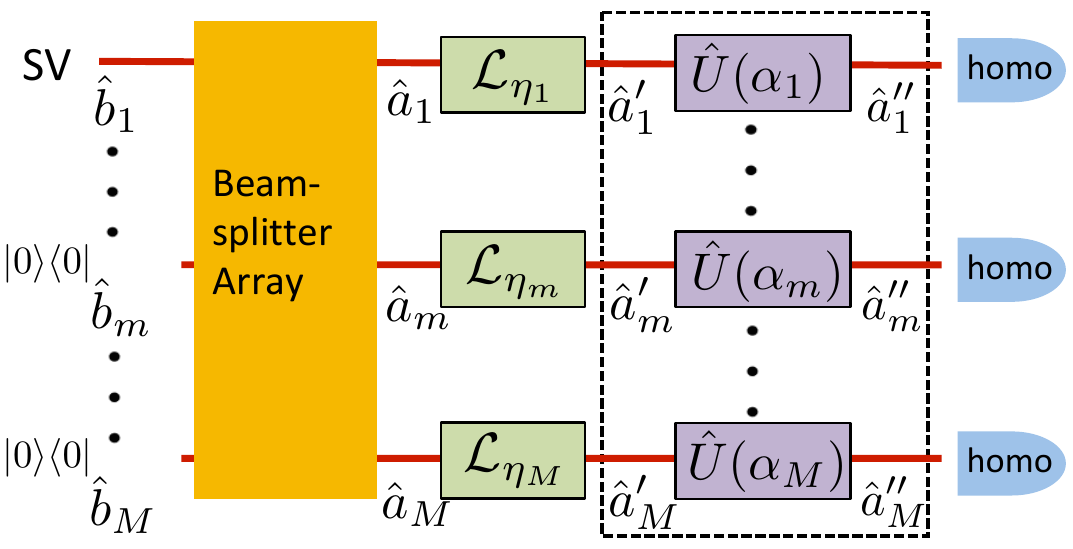}
\caption{Schematic of a distributed quantum sensing protocol for measuring displacements on a single quadrature.  
SV: squeezed-vacuum state with mean photon number $N_S$ and squeezed noise in its real quadrature. $\calL_\eta$:  pure-loss channel with transmissivity $0< \eta \le 1$.  $\hat{U}(\alpha)$:  field-quadrature displacement by real-valued $\alpha$.  homo: homodyne measurement of the real quadrature.   
\label{DS_original_general}
}
\end{figure}

As shown in Fig.~(\ref{DS_original_general}), the original distributed sensing  protocol~\cite{zhuang2018distributed} aims to obtain a minimum rms error estimate of a weighted average, $\bar{\alpha}\equiv \sum_{m=1}^M w_m \alpha_m$, of real quadrature displacements $\{\alpha_m, 1\le m\le M\}$,  where the weights, $\{w_m, 1\le m \le M\}$, are non-negative and sum to one~\footnote{Note that negative weights can be merged into the sign of each $\alpha_m$.}. To do that, in general one inputs $M$ modes, $\hat{a}_m, 1\le m \le M$, one for each sensor node, and performs measurements on the output modes $\hat{a}^{\prime \prime}_m, 1\le m \le M$. To model the imperfections from the distributed sensors, we introduce an independent loss channel $\calL_{\eta_m}$ with transmissivity $\eta_m$ on each sensor node, leading to the mode transform 
$
\hat{a}_m^\prime = \sqrt{\eta_m}\hat{a}_m+\sqrt{1-\eta_m} \hat{e}_m
$, 
where the environment mode $\hat{e}_m$ is in a vacuum state.
To enable performance comparison, we characterize the overall resource for the sensing task by the total mean photon number $N_S$ used in modes $\{\hat{a}_m, 1\le m \le M\}$. This is because for sensing applications like bio-sensing one wants to minimize the light power shining on the fragile samples to avoid any damage.

In an entanglement-enhanced distributed sensing protocol, the modes $\{\hat{a}_m, 1\le m \le M\}$ are in a CV multipartite entangled state, produced by passing a single-mode squeezed vacuum, with mean photon number $N_S$, through a beamsplitter array. Homodyne measurements are applied to obtain the information about the weighted average. To benchmark the performance, we compare the entangled scheme with the optimal separable scheme, where each mode is in a squeezed vacuum state with mean photon number $N_m$, and the total mean photon number $\sum_{m=1}^M N_m=N_S$ for fair comparison. 
In principle, one can introduce extra ancilla modes; however, this is not necessary in the lossless case --- one can show that each scheme is optimal in its own class, given the total mean photon number constraint. In the lossy case, the optimal protocol is still an open question, but refs.~\cite{zhuang2018distributed,xia2018repeater} were able to show that the scheme in Fig.~(\ref{DS_original_general})
maximizes the Fisher information among Gaussian states and achieves the best precision when homodyne measurement is applied. Recently Ref.~\cite{oh2019optimal} proved that this scheme is also the optimal Gaussian protocol for distributed phase sensing.

In the following, we evaluate the performance of the entangled and separable sensing protocols in the presence of loss.
We set the beasmplitters such that $\hat{b}_1 \equiv \sum_{m=1}^Mw_m\sqrt{\eta_m}\,\hat{a}_m/\bar{W}$ is in a squeezed-vacuum state, where $\bar{W} \equiv \sqrt{\sum_{m=1}^M  w_m^2\eta_m}$. Then, we have that $\tilde{\alpha} \equiv \sum_{m=1}^Mw_m{\rm Re}(\hat{a}^{\prime\prime}_m)$ is an unbiased estimator of $\bar{\alpha}$ with the minimum rms error~\cite{zhuang2018distributed},
\be
\delta\alpha_{\bm \eta}^E = \frac{\bar{w}}{2}\left(\frac{\bar{\eta}}{(\sqrt{N_S+1}+\sqrt{N_S})^2}+1-\bar{\eta}\right)^{1/2},
\label{dalpha_extension}
\ee
under the average photon-number constraint, where $\bar{w} \equiv \sqrt{\sum_{m=1}^Mw_m^2}$ and $\bar{\eta} \equiv \sum_{m=1}^M w_m^2\eta_m/\bar{w}^2$.

The optimal separable-state scheme, for the scenario under consideration here, employs a product state, but its precision $\delta \alpha_{\bm \eta}^C$ does not have a closed solution in general.
For the simple case of $\eta_m=\eta, w_m=1/M$, we have~\cite{zhuang2018distributed} 
\begin{align}
&\delta \alpha_\eta^E=\frac{1}{2}\left(\frac{\eta}{M\left(\sqrt{N_S+1}+\sqrt{N_S}\right)^2}+\frac{1-\eta}{M}\right)^{1/2}.
\label{dalpha_E}
\\
&\delta \alpha_\eta^C=\frac{1}{2}\left(\frac{\eta}{M\left(\sqrt{N_S/M+1}+\sqrt{N_S/M}\right)^2}+\frac{1-\eta}{M}\right)^{1/2}. 
\label{dalpha_product}
\end{align}
As shown in Fig.~\ref{QEC_sensing_performance}, in the lossless case, the separable scheme has its precision obeying the SQL, while the entangled scheme achieves the Heisenberg scaling; Even when there is loss, the entanglement enhancement survives, due to the robustness of CV multipartite entanglement. However, the scaling advantage is entirely gone even for small $M$. To mitigate the loss issue, we consider CV quantum error-correction.


\section{Continuous-variable Error correction}
\label{Sec_EC}

Quantum error correction~\cite{calderbank1996} codes are originally developed for protecting DV quantum information for scalable quantum computing, sometimes even with the aid of CV systems~\cite{albert2018performance,cochrane1999macroscopically,gottesman2001}. However, various quantum sensing applications require CV quantum information processing. To facilitate these applications, the question to be addressed in this section is: can we protect continuous-variable quantum information against noise?

The general idea to correct a CV mode is to encode a single mode into multiple modes. Indeed, previous such proposals can correct single-mode errors~\cite{lloyd1998,braunstein1998error}. However, a key difference of CV systems is that errors (e.g., thermal noises and excitation loss) happen with {\em unity} probability on {\em all} modes.
Thermal noise can be described by an additive white Gaussian noise channel (AWGN) $\Phi_{\sigma^2}$, which applies a Gaussian distributed complex-valued random displacement
$
\hat{a}^\prime=\hat{a}+(\epsilon_q+i\epsilon_p)/\sqrt{2}
$
on the input mode.
Here $\epsilon_p,\epsilon_q$ are real Gaussian distributed with standard deviation $\sigma$. In fact, it suffices to consider AWGN channels for all Gaussian noise models, due to channel reduction relations~\cite{caruso2006one,garcia2012majorization,Weedbrook_2012,noh2018quantum,rosati2018narrow}. As an example, the excitation loss channel $\calL_\eta$ can be combined with an amplification channel $\calA_{G}$ in front, described by the mode transform $\hat{a}^\prime=\sqrt{G}\hat{a}-\sqrt{G-1}\hat{e}^\dagger$ joint on the vacuum environment mode $\hat{e}$. Choosing the gain $G=1/\eta$, we 
obtain the composite channel $\calL_\eta\circ \calA_{1/\eta}=\Phi_{1-\eta}$~\footnote{Because $\calA_{1/\eta}\circ
  \calL_\eta=\Phi_{1/\eta-1}$ and $\calL_\eta\circ
  \calA_{1/\eta}=\Phi_{1-\eta}$ and $1-\eta\le 1/\eta-1$, it is always
  beneficial to apply amplification before the loss.}.

\begin{figure}
\centering
\includegraphics[width=0.4\textwidth]{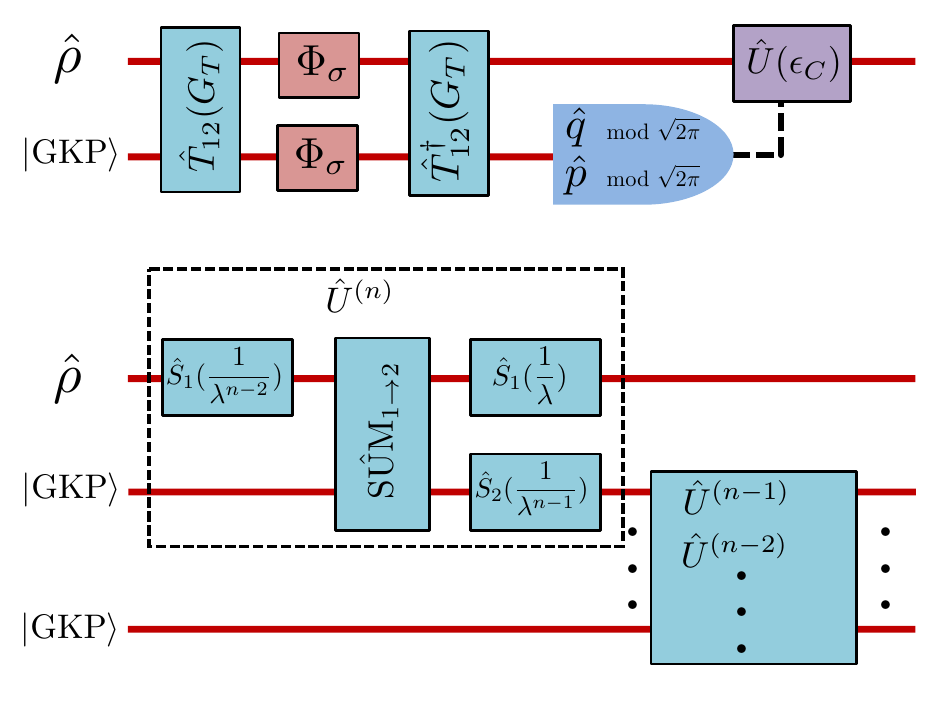}
\caption{Schematic of the GKP-two-mode-squeezing code (top) and the encoding part of the GKP-stabilizer code (bottom).
\label{QEC_schematic}
}
\end{figure}

To correct the AWGN noise, the new idea in Ref.~\cite{noh2019encoding} is to use GKP grid states to encode CV information. This builds on an observation emphasized in ref.~\cite{terhal2016encoding,duivenvoorden2017} --- that with GKP grid states we can simultaneously measure two quadratures with high precision, as long as we're promised that the displacement of both quadratures is small. The GKP grid state has wave function
\begin{align}
\ket{{\rm GKP}}\propto
\sum_{t=-\infty}^\infty e^{-\pi \Delta^2 t^2 }
\int e^{-(q-\sqrt{2\pi}t)^2/2\Delta^2} \ket{q} dq
\nonumber
\\
\propto 
\sum_{t=-\infty}^\infty  \int e^{-\Delta^2p^2/2} e^{-(p-\sqrt{2\pi}t)^2/2\Delta^2}\ket{p}dp.
\end{align}
When $\Delta\ll1$, its Wigner function is peaked around a square grid of spacing $\sqrt{2\pi}$ in the phase space. The overall variance $\braket{\hat{q}^2}\simeq \braket{\hat{p}^2}\simeq 1/2\Delta^2$ equals the mean photon number $N_S$; however, if we consider only the phase space region close to a single peak, the variances in position and momentum are $\Delta^2/2\simeq 1/4N_S\ll 1$, only twice the squeezed-vacuum variance. 



Below we will recall two codes introduced in Ref.~\cite{noh2019encoding}, the GKP-two-mode-squeezing code and the GKP-stabilizer code. 
To understand the error correction mechanism, consider two input modes to an encoding circuit. Mode 1 will be used to detect the signal, and mode 2 is an ancilla which has been prepared in a GKP state. The encoding circuit applies a Gaussian unitary operator $\hat U_{\bm S}$ to this pair of modes, where $\bm S$ is a symplectic tranformation; its inverse $\hat U_{\bm S}^\dagger$ can be used to decode the state. Between encoding and decoding, the two modes are subjected to additive noise --- the noise operation is a displacement $\hat U({\bm \epsilon})= \hat U_1(\epsilon_1)\otimes \hat U_2(\epsilon_2)$. Thus, after decoding, the noise is transformed to a modified displacement $\hat{U}_{\bm S}^\dagger \hat{U}(\bm \epsilon)\hat{U}_{\bm S}=\hat{U}(\bm \epsilon^\prime)$, where $\bm \epsilon^\prime={\bm S}^{-1}\bm \epsilon$~\cite{zhuang2019scrambling}.
By choosing a proper entangling transform $\hat{U}_{\bm S}$, one induces a correlation of the effective displacements $\epsilon_1^\prime$ and $\epsilon_2^\prime$ of the two modes. Then by measuring the displacement $\epsilon_2^\prime$ of the GKP ancilla, one can infer a displacement $\hat{U}(\epsilon_c)$ which corrects the additive error $\epsilon_1^\prime$ on the signal mode. In this scheme, while all operations are Gaussian, the input ancilla is a non-Gaussian GKP grid states, so the effectiveness of error correction is compatible with the no-go theorem for Gaussian error correction in ref.~\cite{niset2009nogo}.

\subsection{GKP-two-mode-squeezing code}
\label{Sec_EC_tms}
\begin{figure}
\centering
\subfigure{
\includegraphics[width=0.225\textwidth]{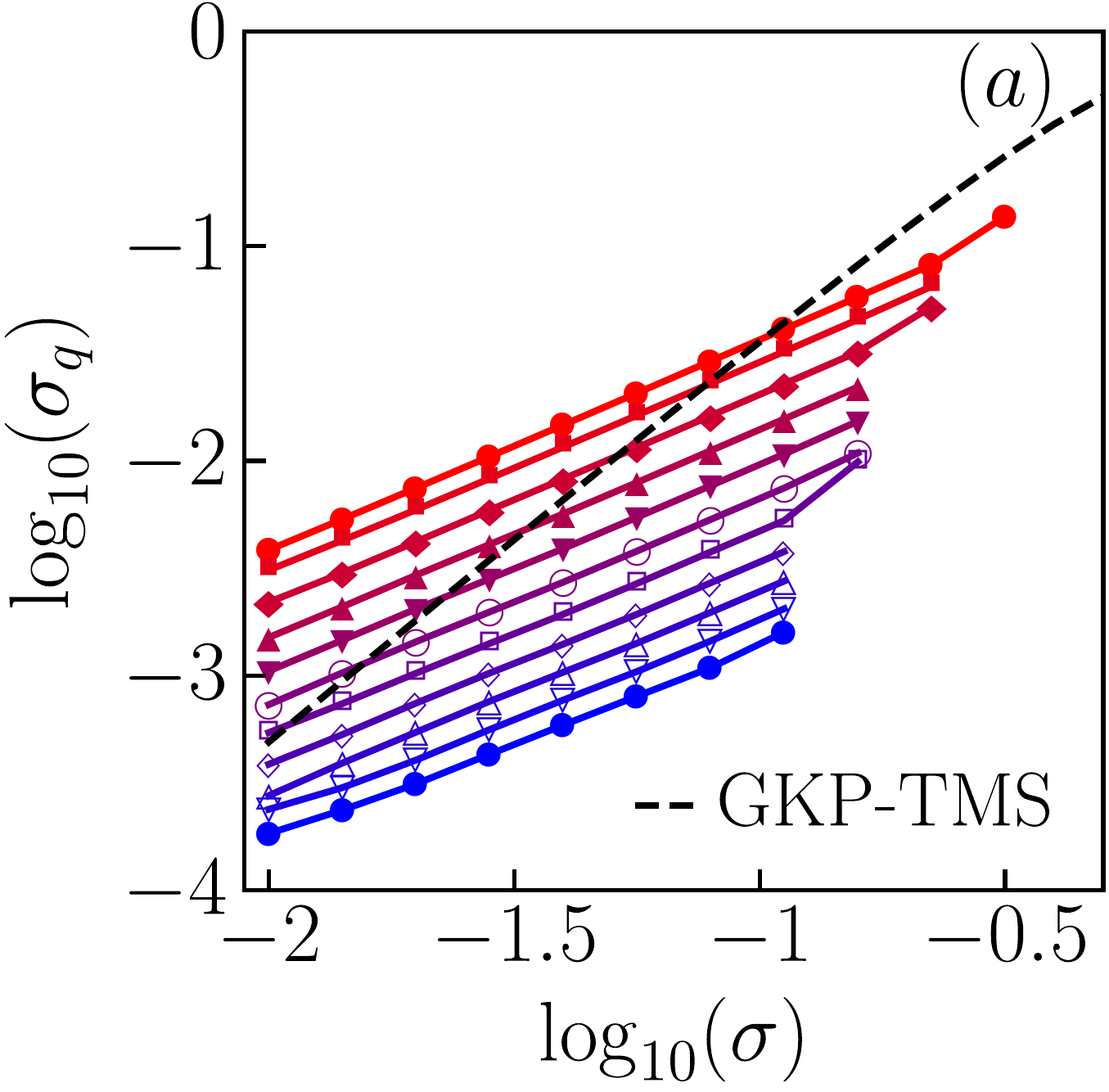}
}
\subfigure{
\includegraphics[width=0.225\textwidth]{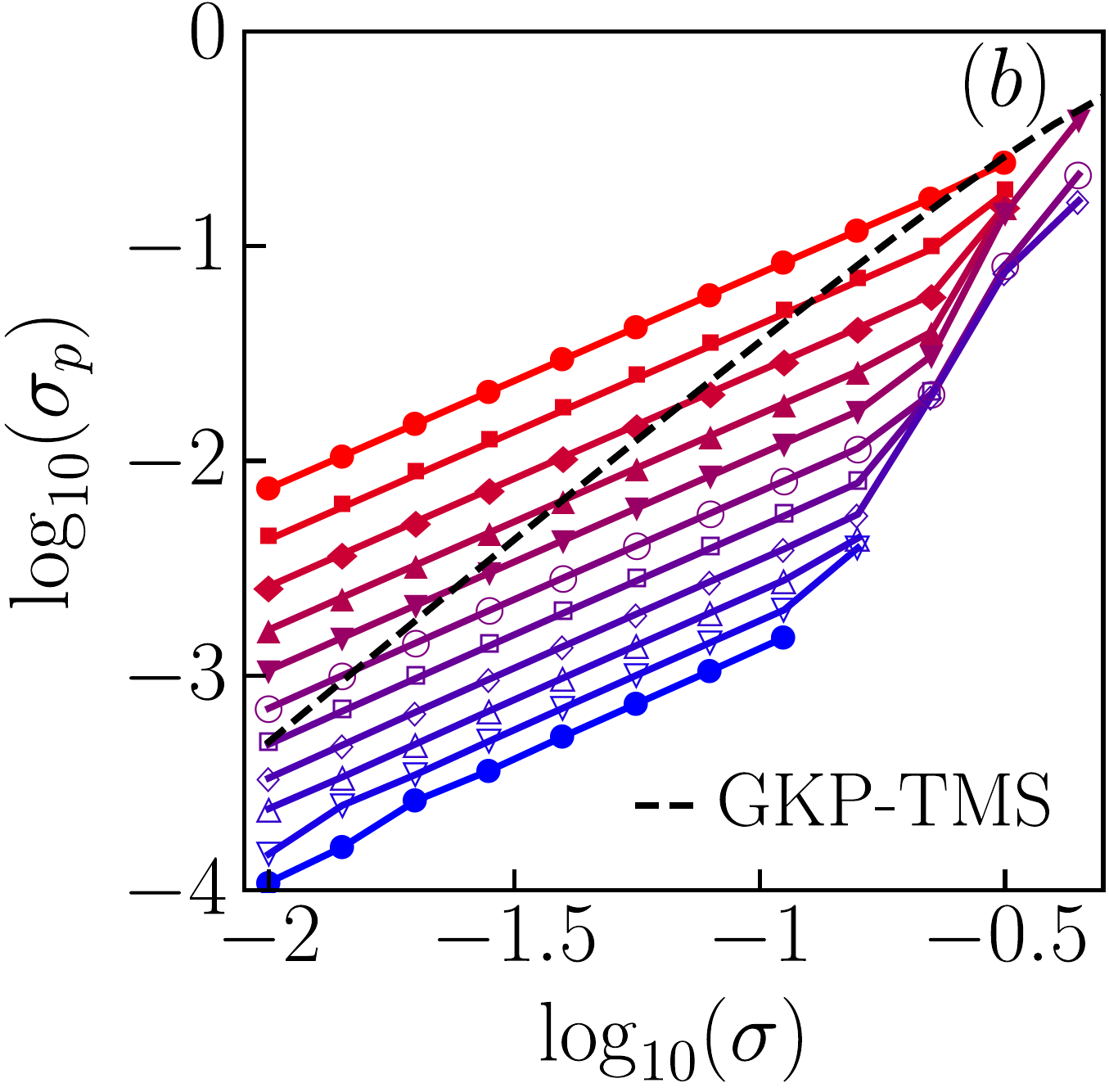}
}
\caption{Performance of the GKP-two-mode-squeezing code and the GKP-stabilizer code. (a) position-quadrature logical noise $\sigma_q$ vs the physical noise $\sigma$. (b) momentum-quadrature logical noise $\sigma_p$ vs the physical noise $\sigma$. The colors from red to blue indicates the level of squeezing from $\lambda=1.05,1.15,\cdots,1.95,2.05$ in the GKP-stabilizer codes.
\label{QEC_performance}
}
\end{figure}

As illustrated in Fig.~\ref{QEC_schematic}, the GKP-two-mode-squeezing code uses a two-mode squeezing operation $\hat{T}_{12}(G_T)$ to entangle the input state with an ancilla initialized in the GKP grid state. After both modes go through the noise channel $\Phi_\sigma$, another conjugate two-mode squeezing operation $\hat{T}_{12}^\dagger(G_T)$ is performed. Finally, both quadratures of the ancilla are measured modulo $\sqrt{2\pi}$ to diagnose the displacement error on the input state. After conjugation of the displacement error by the two-mode squeezing operator, we obtain the the effective displacements $\hat{T}_{12}^\dagger(G_T) \left[\hat{U}_1(\epsilon_1)\otimes \hat{U}_2(\epsilon_2) \right] \hat{T}_{12}(G_T)=\hat{U}_1(\sqrt{G_T}\epsilon_1+\sqrt{G_T-1}\epsilon_2)\otimes \hat{U}_2(\sqrt{G_T}\epsilon_2+\sqrt{G_T-1}\epsilon_1)$; we see that when $G_T$ is large, the effective displacements are highly correlated. Thus measuring the displacement noise of the ancilla provides a good estimate of the displacement error on the signal, and therefore enables approximate correction of the error through a counter-displacement. Although the uncertainty principle forbids the simultaneous precise measurement of displacements on both quadratures, an ancilla in the GKP grid state allows the precise measurement of both quadrature displacements modulo $\sqrt{2\pi}$. Ref.~\cite{noh2019encoding} has given a detailed analysis on the amount of noise reduction in this scheme. We plot the rms logical noises $\sigma_p,\sigma_q$ given by Eq. 24 in Ref.~\cite{noh2019encoding} on each quadrature with the physical noise $\sigma$ in Fig.~\ref{QEC_performance}. The code helps when $\sigma\leq 0.558$, which corresponds to loss $\eta\geq 0.689$.

\subsection{GKP-stabilizer codes}
\label{Sec_EC_stabilizer}

Ref.~\cite{noh2019encoding} has also proposed a more general GKP-stabilizer code. In the GKP stabilizer code, a hierarchical structure of squeezing and GKP encoding is used, as shown in Fig.~\ref{QEC_schematic}. We begin by analyzing the lowest level ($n=2$). 

The encoding is achieved by a sequence of Gaussian operations --- a two-mode SUM gate and single-mode squeezing operations.
Recall that a SUM gate $\hat{\rm SUM}_{1\to 2}$ acts on a pair of modes according to
\begin{equation}
\hat{{\rm SUM}}_{1\to 2}: (\hat{q}_1, \hat{p}_1, \hat{q}_2, \hat{p}_2) \to (\hat{q}_1, \hat{p}_1 - \hat{p}_2, \hat{q}_1 + \hat{q}_2, \hat{p}_2),
\end{equation}
and that a squeezing operation acts on mode $m$ according to
\begin{equation}
\hat{S}_m(\lambda): (\hat{q}_m,\hat{p}_m) \to (\lambda\hat{q}_m, \hat{p}_m/\lambda).
\end{equation}
To protect mode 1, we make use of the GKP ancilla mode 2 via the encoding circuit
\begin{equation}
\hat{U}^{(2)}= \hat{S}_1(\gamma)\hat{S}_2(\beta) \hat{{\rm SUM}}_{1\to 2}\hat{S}_1 (\delta).
\end{equation}
Here the order of operations is read from right to left --- that is, $\hat{S}_1(\delta)$ acts first, followed by the sum gate and then $\hat{S}_1(\gamma)\hat{S}_2(\beta)$. The overall $4\times 4$ symplectic matrix applied to the four quadratures can be calculated to be
\begin{equation}
S_{\hat{U}^{(2)}} = \left(
\begin{matrix}
\delta\gamma & 0 & 0 & 0\\
0 & \frac{1}{\delta \gamma} & 0 & -\frac{1}{\gamma}\\
\delta\beta & 0 & \beta & 0 \\
0 & 0 & 0 & \frac{1}{\beta}
\end{matrix}\right).
\end{equation}
The decoding circuit is the encoder run in reverse and implements the unitary $\hat{U}^{(2)\dagger}$.

Now suppose that after encoding, additive noise acts on the two modes leading to the displacement $\bm \epsilon=(\epsilon_1^q,\epsilon_1^p,\epsilon_2^q,\epsilon_2^p)^T$
and then the decoder is applied. If there is no noise, the decoder perfectly restores the input signal. But when there is noise the decoder distorts the noise, yielding noise in the output signal
\be
\bm \epsilon^\prime=S_{\hat{U}^{(2)}}^{-1}\bm \epsilon = 
\left(\frac{1}{\delta\gamma}\epsilon_1^q, \delta\gamma\epsilon_1^p +\delta\beta \epsilon_2^p,-\frac{1}{\gamma} \epsilon_1^q + \frac{1}{\beta} \epsilon_2^q, \beta\epsilon_2^p\right)^T.
\ee
Because mode 2 was initially encoded as an ideal GKP grid state, it is possible to simultaneously measure the offset of both quadratures in mode 2, assuming the (distorted) noise is sufficiently weak. Once we know the offset in mode 2, we can approximately diagonose the additive shift in the (unmeasured) mode 1. Specifically, after measuring
\begin{equation}
\Delta q_2 = -\frac{1}{\gamma} \epsilon_1^q + \frac{1}{\beta} \epsilon_2^q, \quad 
\Delta p_2 = \beta\epsilon_2^p,
\end{equation}
we apply a corrective displacement $\bm \epsilon_c=( \Delta q_2/\delta, -\delta \Delta p_2)^T$,
obtaining the partially corrected noise in mode 1:
\begin{equation}
\bm \epsilon_1^{\prime\prime}=\left(\frac{1}{\delta\beta}\epsilon_2^q,\delta\gamma \epsilon_1^p\right)^T.
\end{equation}

Suppose for example that $\delta = 1$, $\beta = \gamma^{-1} = \lambda > 1$, and that $\epsilon_1^q$, $\epsilon_1^p$, $\epsilon_2^q$, $\epsilon_2^p\approx\epsilon$ are all comparable; then the noise in both quadratures of mode 1 is suppressed by a factor $\lambda^{-1}$ relative to an unprotected mode. In practice the noise suppression is limited because, for a given noise strength there is a limit to how much we can squeeze the noise and still read out both quadratures of the GKP grid state unambiguously. Furthermore, if the GKP grid states themselves are only finitely squeezed, further squeezing during the protocol may compromise their error-correcting power. 

We can go further in a protocol that makes use of multiple GKP-encoded modes. To see how that works, we consider the case where the noise acting on the two modes is asymmetric, so that $\epsilon_2^q \approx \epsilon_2^p\approx \epsilon_2 < \epsilon_1 \approx \epsilon_1^q \approx \epsilon_1^p$, and we adjust the protocol so that the output noise on mode 1 after decoding and recovery is balanced between $q$ and $p$; hence
\begin{equation}
\delta\gamma\epsilon_1 = \frac{1}{\delta \beta} \epsilon_2 \implies \kappa = \delta^2\beta\gamma, \quad {\rm where} \quad \kappa = \epsilon_2/\epsilon_1<1.
\end{equation}
There is a further constraint --- we don't want the shift error in mode 2 after decoding to be too large, which would compromise our ability to measure both quadratures accurately. To ensure that the distorted error is comparable in both quadratures in mode 2, we impose
\begin{equation}
\frac{1}{\gamma} \epsilon_1 = \beta\epsilon_2 =\beta\kappa\epsilon_1\implies \kappa = \frac{1}{\beta\gamma}\implies \delta=\kappa.
\end{equation}
To summarize, if the noise in mode 2 is weaker than the noise in mode 1 by the factor $\kappa < 1$, and if we want the error-corrected noise in mode 1 to be balanced between the $q$ and $p$ quadratures, we use the encoder with
\begin{equation}
\gamma = \frac{1}{\lambda},\quad \beta = \frac{\lambda}{\kappa},\quad \delta = \kappa < 1,
\end{equation}
where $\lambda > 1$; then the error corrected noise in mode 1 is $\epsilon_2/\lambda$ in both quadratures. 

Because the scheme works for asymmetric noise, it can be used iteratively. For example, with three modes, where mode 1 is the sensing mode and modes 2 and 3 are GKP-encoded ancillas, we can use mode 3 to reduce the additive noise in mode 2 by a factor of $1/\lambda$, and then use mode 2 to reduce the noise in mode 1 by a further factor of $1/\lambda$, achieving all together a reduction by $1/\lambda^2$ in the noise in the sensing mode. If there are $n{-}1$ GKP-encoded ancillas, then as indicated in Fig.~\ref{QEC_schematic} the encoding circuit acting on modes 1 and 2 would be 
\be 
\hat{U}^{(n)}= \hat{S}_1(\frac{1}{\lambda}) \hat{S}_2(\lambda^{n-1})\hat{{\rm SUM}}_{1\to2}
\hat{S}_1(\frac{1}{\lambda^{n-2}}),
\ee 
which implements the symplectic transformation
\be 
S_{\hat{U}^{(n)}}
=\begin{pmatrix} 
\lambda^{1-n} & 0 & 0 & 0\\
0 & \lambda^{n-1} & 0 & -\lambda\\
\lambda & 0 & \lambda^{n-1} & 0\\
0 & 0 & 0 & \lambda^{1-n}
\end{pmatrix}.
\ee 

Guided by this intuition, let's formally optimize the decoding operations. Suppose on the $(n-1)$-th level the logical noises $Z_{q,n-1}$ and $Z_{p,n-1}$ have been reduced to $\sigma_{q,n-1}$ and $\sigma_{p,n-1}$; then the initial covariance matrix of the additive noise on the ancilla $Z_{q,n}^{(0)}, Z_{p,n}^{(0)}$ and the $(n-1)$-th level logical noise is $V_{0,n-1}={\rm Diagonal}[\sigma^2,\sigma^2, \sigma_{q,n-1}^2, \sigma_{p,n-1}^2]$. After the encoding and decoding the effective noise covariance matrix
$V_{ED,n}=S_{\rm \hat{U}^{(n)\dagger}} V_{0,n-1}S_{\rm \hat{U}^{(n)\dagger}}^T 
$.
The decoding operation measures the ancilla of the $(n-1)$-th level and performs a displacement on the $n$-th level ancilla.
\begin{align}
Z_{q,n}= Z_{q,n}^{(0)}-C_qR_{\sqrt{2\pi}}(Z_{q,n-1}),  
\\
Z_{p,n}= Z_{q,n}^{(0)}-C_pR_{\sqrt{2\pi}}(Z_{p,n-1}).
\end{align}
Here $R_s(z)=z-n^\star(z)s$ and $n^\star(z)=\argmin_{n\in \mathbb{Z}}|z-ns|$, i.e. function $R_s(z)$ takes the generalized modulo $z\mod s$. The choice of the coefficients minimizes the variance in the ideal case,
\be
C_q=\frac{V_{ED,n}(1,3)}{V_{ED,n}(3,3)}, C_p=\frac{V_{ED,n}(2,4)}{V_{ED,n}(4,4)}.
\ee
Due to the imperfect measurement of quadratures that a GKP grid state offers, after the error correction the probability density function (PDF) of the logical noise $P_{Z_{x,n}}(\cdot), x=p,q$ is not Gaussian. It can be obtained through the following recursion relation,
\begin{align}
&P_{Z_{x,n}}(\xi_x)=
\int d \xi_0 \int d \xi_1 P_{Z_{x,n}^{(0)}}(\xi_0) P_{Z_{x,n-1}}(\xi_1)
\nonumber
\\
&
\ \ \ \ \ \ \ \ \ \ \ \ \ \ \times \delta[\xi_x-\xi_0-C_x R_{\sqrt{2\pi}}(\xi_1)]
\nonumber
\\
&= 
\sum_{n=-\infty}^\infty \int_{n_-\sqrt{2\pi}}^{n_+\sqrt{2\pi}} d \xi_1 P_{Z_{x,n}^{(0)}}[\xi_x-C_x R_{\sqrt{2\pi}}(\xi_1)] P_{Z_{x,n-1}}(\xi_1),
\label{int_transform}
\end{align}
where $n_\pm=n\pm 1/2$.
Note that $P_{Z_{x,n}^{(0)}}(\cdot)$ and the initial noise $P_{Z_{x,1}}(\cdot)$ both obey zero-mean Gaussian distributions with variance $\sigma^2$. The variances of the logical noises can be obtained by integrations over the PDFs obtained from the recursion.

Suppose all measurements are perfect (without the $\sqrt{2\pi}$ ambiguity), then the ideal evolution of noise obeys
\be
\sigma_{q,n}^2\simeq \frac{\sigma_{q,n-1}^2}{\lambda^2(1+\lambda^{-2n}\sigma^{-2}\sigma_{q,n-1}^2)},
\sigma_{p,n}^2\simeq \lambda^{2-2n} \sigma^2.
\ee
In the small noise limit, we have $\sigma_{q,n}^2\simeq \sigma_{p,n}^2\simeq \lambda^{2-2n} \sigma^2$, which indeed agrees with the intuitive understanding.

To summarize, if there is one input signal model we wish to protect in a sensing experiment, we may introduce $n{-}1$ ancilla GKP grid states and then iterate the protocol $n{-}1$ times, thereby reducing the noise strength to $\sigma_{n-1} \simeq \sigma/ \lambda^{n-1}$ in the error-corrected signal state. To precisely evaluate the error correction performance under moderate noise, we perform numerical integration of Eq.~\ref{int_transform} repeatedly and obtain the standard deviation $\sigma_{q,n}, \sigma_{p,n}$. 
The results, for the case $n=7$, are in Fig.~\ref{QEC_performance}, where different levels of the squeeze parameter $\lambda$ are chosen, indicated by the color; we did not perform the computations for larger values of $n$ due to limitations on numerical precision.
When the initial noise $\sigma$ is large, the modulo $\sqrt{2\pi}$ property of the measurement leads to excess noise and thus hinders the error correction performance. At certain critical noise level, the code ceases to reduce the noise, as indicated by the termination of the plots on the right-hand-side in Fig.~\ref{QEC_performance}.  

\section{Improved distributed sensing}
\label{Sec_improvement}

Now we apply the CV error correction codes in the distributed sensing protocol introduced in section~\ref{Sec_distributed_sensing}. We will evaluate the standard deviations in parameter estimation given the loss and error correction. Although the GKP based error correction codes lead to non-Gaussian random errors in the parameter estimation, the standard deviation of noise is still a good characterization of the measurement precision. This is because in a parameter estimation scenario, one can average multiple independent repetitions of the same measurement, even when each measurement is already multi-mode. When the number of repetitions is large, the central limit theorem guarantees that the averaged measurement error is Gaussian distributed and thus can be characterized entirely by its standard deviation.

\subsection{Error corrected real quadrature sensing}
\label{Sec_improvement_real}

\begin{figure}
\centering
\includegraphics[width=0.45\textwidth]{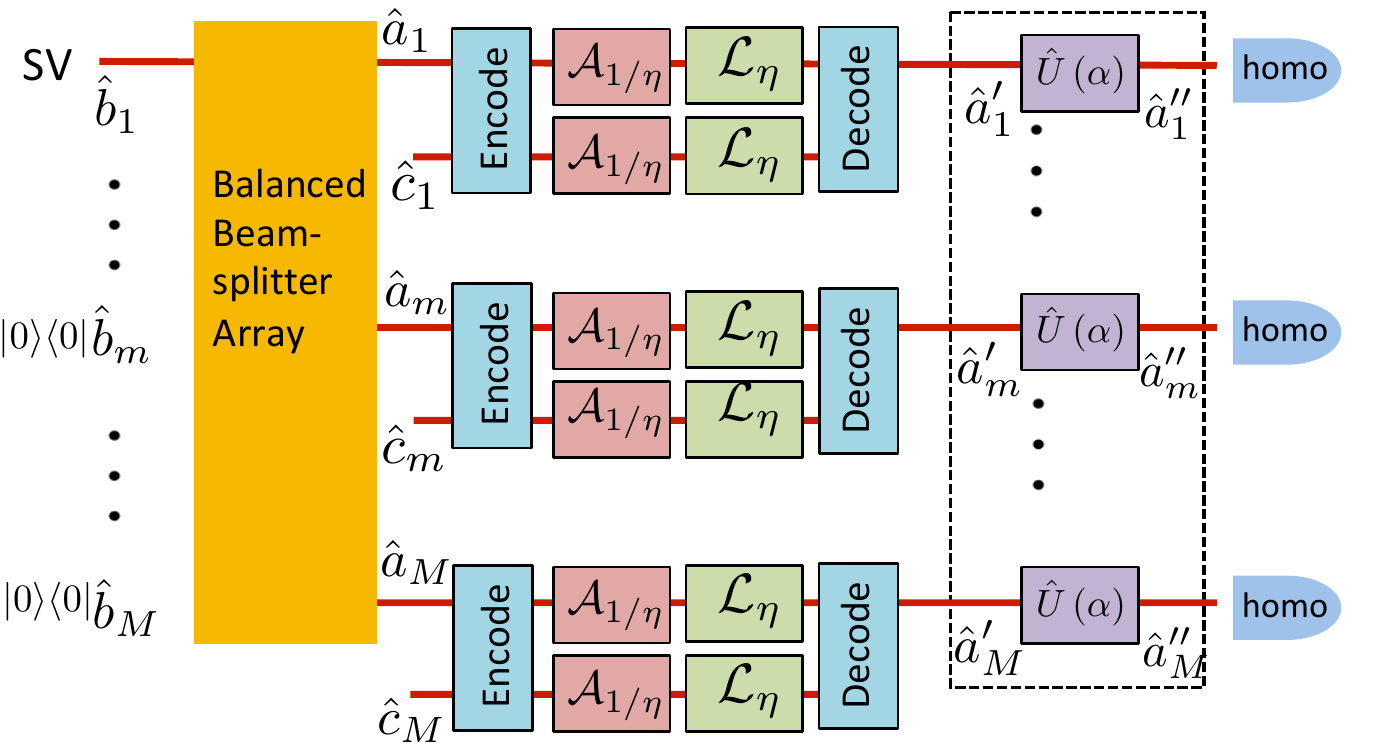}
\caption{Distributed quantum sensor for measuring field-quadrature displacement with error correction in the distribution step. Multiple anculla modes can be utilized, however, only a single ancilla mode is shown for simplicity.
\label{DS_original_general_correct}
}
\end{figure}

We apply the GKP-two-mode-squeezing code and the GKP-stabilizer code in a real-quadrature distributed sensing protocol. As shown in Fig.~\ref{DS_original_general_correct}, to perform distributed sensing on different nodes, one first locally generates signal modes $\{\hat{a}_m, 1\le m \le M\}$ in the same CV multipartite entangled state as in Section~\ref{Sec_distributed_sensing}. After the beamsplitter array, each mode $\hat{a}_m$ in the multipartite entangled state is immediately encoded (with additional ancilla) to protect against independent loss errors; to facilitate error correction, amplifiers $\calA_{1/\eta}$ transform the loss channels $\calL_\eta$ to AWGN channels that can be corrected with the standard GKP decoder. Before the sensing process, decoding is applied to the received signal modes and ancillae, and then the error-corrected signal inputs, $\{\hat{a}_m^\prime\}$, are injected to sense displacements. Note that, while a total of $\propto M$ ancilla modes are used in the 
entanglement distribution process, the sensing process occurs after the GKP-assisted decoding, and only a single error-corrected signal mode interacts with the sample at each sensing node.

\begin{figure}
\centering
\includegraphics[width=0.45\textwidth]{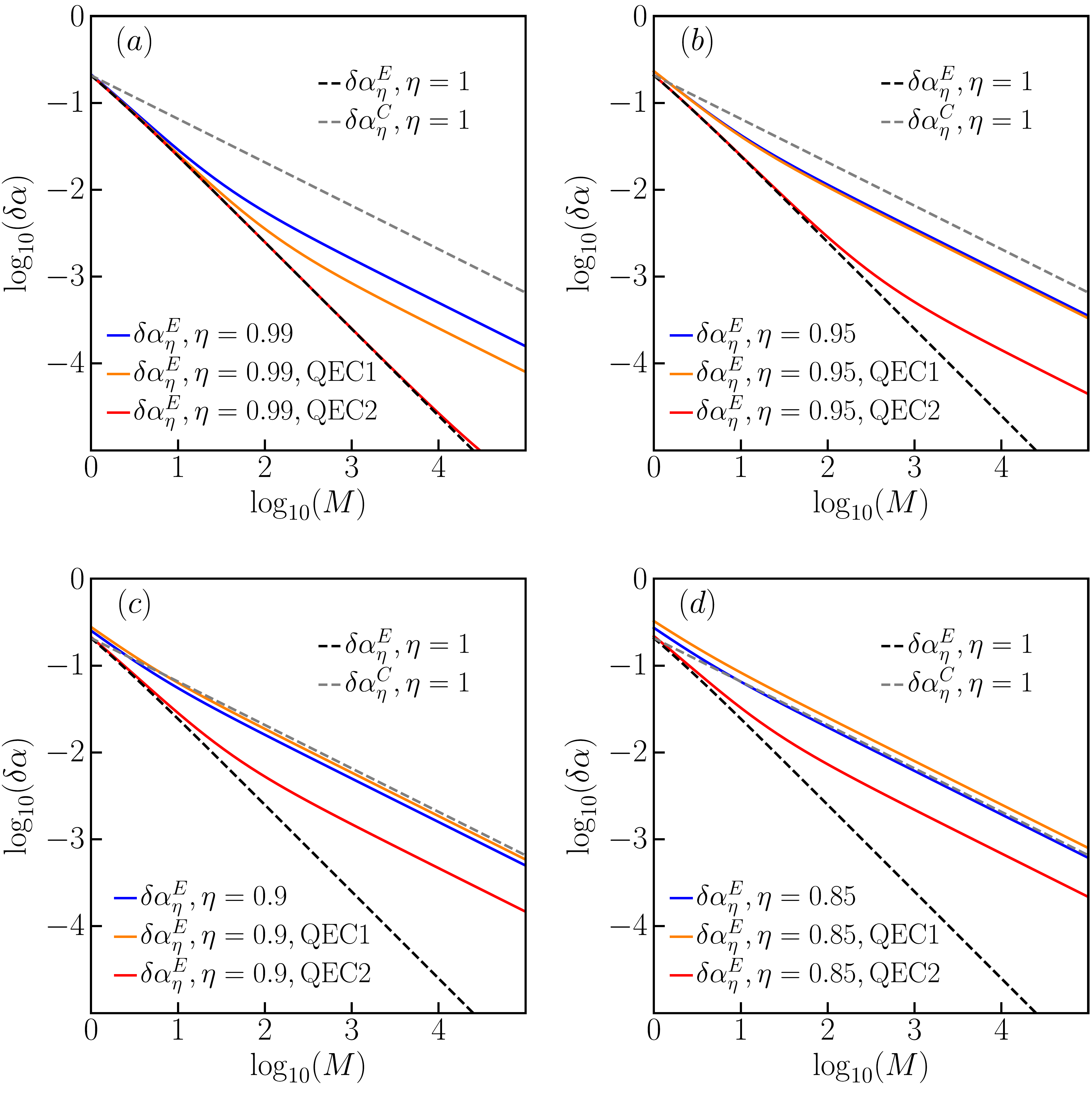}
\caption{
  Performance of error correction codes for distributed real quadrature displacement sensing. $N_S=Mn_S$ with $n_S=1$. $\delta \alpha$ is the measurement standard deviation and $M$ is the number of sensing nodes. While the lossless case is plotted for comparison, we considered various losses $\eta$ (corresponding to noise $\sigma=\sqrt{1-\eta}$ in Fig.~\ref{QEC_performance}) for the protocols: (a) $\eta=0.99$ ($\log_{10}\sigma=-1$), (b) $\eta=0.95$ ($\log_{10}\sigma\simeq-0.65$), (c) $\eta=0.9$ ($\log_{10}\sigma=-0.5$), and (d) $\eta=0.85$ ($\log_{10}\sigma\simeq-0.41$). QEC1: GKP-two-mode-squeezing code. QEC2: GKP-stabilizer code with $n=7$. Note that in (c) and (d) the performance of QEC 1 is worse than non-corrected case, due to the extra amplification required to reduce loss to additive thermal noise. 
  \label{QEC_sensing_performance}
}
\end{figure}

As a demonstration, we consider the case of equal weights and equal displacements; similar advantages are expected for more generic cases. Suppose the error correction code reduces the original noise $\sqrt{1-\eta}$ to $\sigma_{\rm EC}(\eta)$; as in Eq.~\ref{dalpha_E}, one can then obtain the precision
\be 
\delta \epsilon_\eta^{EC}=\frac{1}{2}\left(\frac{1}{M\left(\sqrt{N_S+1}+\sqrt{N_S}\right)^2}+\frac{2\sigma_{EC}^2(\eta)}{M}\right)^{1/2}.
\label{dalpha_EC}
\ee 
Note that, due to the amplification, compared with Eq.~\ref{dalpha_E} the error is larger by a factor $\sqrt{\eta}$, and also the second term inside the square root is a factor of two larger in addition to the change from $1-\eta$ to $\sigma_{\rm EC}^2(\eta)$.
Using the results in section~\ref{Sec_EC}, we can evaluate the performance of both error correction codes. As shown in Fig.~\ref{QEC_sensing_performance}, we see that the GKP-two-mode-squeezing code (orange) only has a small advantage over the scheme without error correction (blue) in the low loss region ($\eta\leq 0.95$). The GKP-stabilizer code with $n=7$ (red) gives a much better performance improvement. 
In the low loss region , the Heisenberg scaling of of precision can be reinstated up to $M\sim 10^2$ modes. Moreover, when $\eta=0.85$ there is still appreciable advantage over the scheme without error correction (blue). 

A few comments are worthy of mention here. First, the above performance is valid for arbitrary displacement values, but we can do better the if displacement at each sensing node is guaranteed to be smaller than $\sqrt{2\pi}$. A GKP-decoding error could result in a displacement of $\hat a'$ by an unknown integer multiple of $\sqrt{2\pi}$, but this error has no damaging effect if we decode the result of the homodyne measurement of each mode by evaluating it modulo $\sqrt{2\pi}$.
Second, in a fair comparison between sensing schemes, we usually fix the mean photon number of the source that interacts with the samples. In the above comparison, we have not quite done that, because in the case without error correction the mean photon number at the sensing nodes has been attenuated by the loss factor $\eta$, while in the case with error correction we have compensated for the loss channel $\mathcal{L}_\eta$ with the amplification channel $\mathcal{A}_{1/\eta}$, which transforms the loss channel into an AWGN channel. 
However, as one can see in Fig.~\ref{QEC_sensing_performance}, we have plotted the separable scheme in the lossless case (gray dashed) for comparison---it has the same input mean photon number $\sim N_S/M$, while its performance is limited by the SQL. Also, the dominant noise in the entangled scheme without error correction (blue) comes from loss, and further increasing the initial mean photon number $N_S$ barely changes the performance.
Finally, we address the necessity of the GKP grid states. For single-quadrature measurement, one might think that a CV repetition code~\cite{lloyd1998} will also be able to suppress the noise due to an effective squeezing. However, in that case while the noise in one quadrature decreases, the noise in the other quadrature increases, leading to an overall increase in the mean photon number. Only a code with non-Gaussian resources such as GKP grid states can suppress noise in both quadratures.

\subsection{Distributed sensing for complex-valued displacements}
\label{Sec_improvement_complex}
\begin{figure}[t]
\centering
\includegraphics[width=0.4\textwidth]{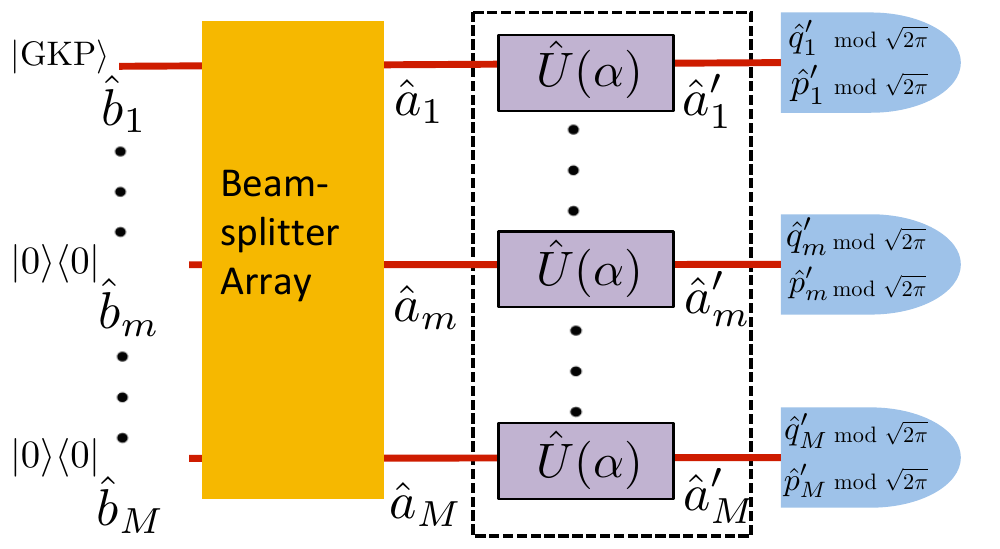}
\caption{Distributed quantum sensor for measuring complex-valued displacements.
\label{DS_GKP_no_loss}
}
\end{figure}

The original distributed sensing protocol~\cite{zhuang2018distributed} only estimates displacements on a single quadrature.
Because a GKP grid state enables the precise measurement of both quadratures~\cite{terhal2016encoding,duivenvoorden2017} when the displacements are small, we hope to utilize GKP grid states to extend the distributed sensing protocol to estimate complex-valued displacements.
Let's start with the simple lossless and equal-weight case, later we will address the extensions to the unequal-weight and lossy cases. Compared with the original protocol in Fig.~\ref{DS_original_general}, the new protocol (schematic in Fig.~\ref{DS_GKP_no_loss}) has mode $\hat{b}_1$ in a GKP grid state instead of a squeezed vacuum state; The output modes from passing the mode $\hat{b}_1$ through the same beamsplitter array go through displacements $\hat{U}(\alpha)$, with complex $\alpha$. Finally one utilizes simultaneous measurement of quadratures with a modulo $\sqrt{2\pi}$ constraint. 

Below we design estimators for both the real and imaginary parts of $\alpha$ and analyze their performance.
A direct attempt to create estimators of ${\rm Re}(\alpha)$ and ${\rm Im}(\alpha)$ is simply to take the averages of the measurement results, $\{\widetilde{\hat{q}_m^\prime}=R_{\sqrt{2\pi}}(\hat{q}_m^\prime), 1\le m \le M\}$ and $\{\widetilde{\hat{p}_m^\prime}=R_{\sqrt{2\pi}}(\hat{p}_m^\prime), 1\le m \le M\}$. However, because we have distributed a single GKP grid state to multiple nodes, there will be large vacuum fluctuations on each node, causing modulo $\sqrt{2\pi}$ errors even when $\alpha$ is arbitrarily small. To avoid those errors, we consider the estimators
\begin{align}
&\sqrt{2}\widetilde{{\rm Re}(\alpha)} =R_{\frac{\sqrt{2\pi}}{M}}\left(\frac{1}{M}\sum_m \widetilde{\hat{q}_m^\prime}\right) =R_{\frac{\sqrt{2\pi}}{M}}\left(\frac{1}{M}\sum_m \hat{q}_m^\prime\right),
\nonumber
\\
&\sqrt{2}\widetilde{{\rm Im}(\alpha)}=R_{\frac{\sqrt{2\pi}}{M}}\left(\frac{1}{M}\sum_m \widetilde{\hat{p}_m^\prime}\right) =R_{\frac{\sqrt{2\pi}}{M}}\left(\frac{1}{M}\sum_m \hat{p}_m^\prime\right).
\label{estimator_complex}
\end{align}
The expectation values of the above estimators are $\braket{\sqrt{2}\widetilde{{\rm Re}(\alpha)}}=R_{\frac{\sqrt{2\pi}}{M}}\left(\sqrt{2}{\rm Re}(\alpha)\right)$ and $\braket{\sqrt{2}\widetilde{{\rm Im}(\alpha)}}=R_{\frac{\sqrt{2\pi}}{M}}\left(\sqrt{2}{\rm Im}(\alpha) \right)$. Although the estimators give the displacement values with a modulo degeneracy, when a good prior knowledge of $\alpha$ is available, the modulo will not introduce additional noise. In the following, we make this heuristic rigorous.

Because of the symmetry in the GKP grid state and the estimators in Eq.~\ref{estimator_complex}, the performance of the two estimators are identical and we only need to analyze the real quadrature. For the equal-weight scenario, by choosing balanced beamsplitters, we have $\sqrt{2}\widetilde{{\rm Re}(\alpha)}=R_{\frac{\sqrt{2\pi}}{M}}\left(\hat{Q}_1/\sqrt{M}+\sqrt{2}{\rm Re}(\alpha)\right)$, where $\hat{Q}_1=\sqrt{2} {\rm Re}(\hat{b}_1)$ is the position quadrature of the GKP grid state. The distribution of $\hat{Q}_1/\sqrt{M}$ concentrates on points $\{k\sqrt{2\pi}/\sqrt{M},k\in \mathbb{Z}\}$, and around each point there is variance $\Delta^2/2M$. Thus the effect of the modulo errors on the performance of $\sqrt{2}\widetilde{{\rm Re}(\alpha)}$ depends on the choice of $M$ and the amplitude of ${\rm Re}(\alpha)$.

A special case is when ${\rm Re}(\alpha)=0$ and $M$ is a square of an integer, i.e., $M=J^2, J\in \mathbb{Z}$. In this case, the PDF of the estimator $\sqrt{2}\widetilde{{\rm Re}(\alpha)}$ is peaked around zero with variance $\Delta^2/2M$. In general, if $M$ is not a square of an integer, then after the modulo the PDF can concentrate on various peaks across the range of $[-\sqrt{2\pi}/\sqrt{M},\sqrt{2\pi}/\sqrt{M}]$ and cause extra noise. To avoid the extra noises, since $M$ is chosen by design in a sensor network, we can indeed ensure $M$ to be a square of an integer.

However, we still need to address the noise introduced by $\sqrt{2} {\rm Re}(\alpha)$ being nonzero. We argue that when a good prior of $\sqrt{2} {\rm Re}(\alpha)$ is available, the estimator can increase the precision as if $\sqrt{2} {\rm Re}(\alpha)$ is zero. A good prior is often possible when the parameter estimation process has multiple steps. Upon obtaining the prior estimation $q_{\rm prior}$ in the first step, one can apply a displacement $\hat{U}(-q_{\rm prior}/\sqrt{2})$ at each node before the next measurement. Here $q_{\rm prior}$ is the prior estimator's result, which we assume to obey a Gaussian distribution $\calN[\sqrt{2} {\rm Re}(\alpha), \sigma_{\rm prior}^2]$, with mean $\sqrt{2} {\rm Re}(\alpha)$ and variance $\sigma_{\rm prior}^2$. The second measurement result $q_{\rm GKP}$ will obey $\calN[\sqrt{2}{\rm Re}(\alpha)-q_{\rm prior},\Delta^2/2M]$ up to the $\mod \sqrt{2\pi}/M$ constraint. Because of the additional displacement, without loss of generality we can assume ${\rm Re}(\alpha)=0$. To combine the prior with the new measurement result, we construct the new estimator
\be 
\calE_\zeta=(1-\zeta) q_{\rm prior}+\zeta(q_{\rm GKP}+q_{\rm prior})=\zeta q_{\rm GKP}+q_{\rm prior},
\ee
where $\zeta$ will be chosen to minimize the combined variance. 

\begin{figure}
\centering
\includegraphics[width=0.3\textwidth]{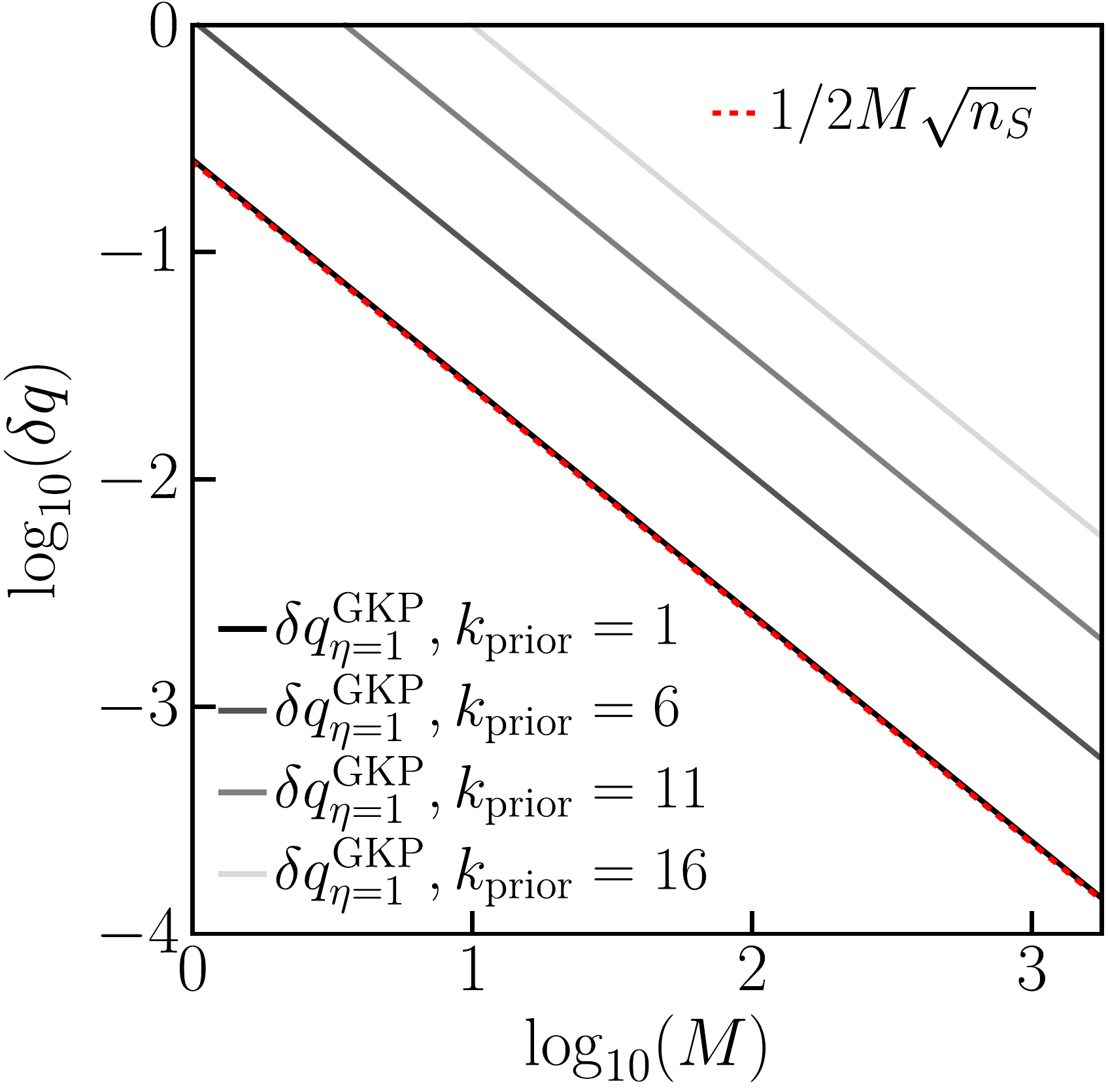}
\caption{The performance of the GKP distributed sensing for complex-valued displacements for various prior information. Without loss of generality, only the precision in the real quadrature displacement is plotted. $n_S=4$.
\label{GKP_M}
}
\end{figure}

Now let's evaluate the performance of this estimator.
First, it is easy to see that the estimator is unbiased, i.e.,
$
\braket{\calE_\zeta}=0.
$
Denote $c_M=\sqrt{2\pi}/M$, $k_\pm=k\pm1/2$, and $\braket{f(q_{\rm prior},x)}_k$ as the average over the prior and measurement outcome in the range of $[k_- c_M,k_+c_M]$, i.e. 
\begin{align}
&
\braket{f(q_{\rm prior},x)}_k=\int d q_{\rm prior} P_{\calN[0, \sigma_{\rm prior}^2]} (q_{\rm prior})
\nonumber
\\
&
\times \int_{k_-c_M}^{k_+c_M} dx\  P_{\calN[-q_{\rm prior},\Delta^2/2M]}(x) f(q_{\rm prior},x).
\end{align}
The overall variance can be obtained through
$
(\delta \calE_\zeta)^2
=\sum_{k=-\infty}^\infty
\braket{[\zeta(x-k c_M)+ q_{\rm prior}]^2}_k
=
\sigma_{\rm prior}^2+\zeta^2 V_1+2\zeta  V_2,
$
where the outcome variance $V_1=\sum_{k=-\infty}^\infty\braket{(x-k c_M)^2}_k$ and the cross correlation $V_2=\sum_{k=-\infty}^\infty\braket{(x-k c_M)q_{\rm prior}}_k$.
It is easy to see that the minimum variance, achieved at $\zeta^\star=-V_2/V_1$, equals
\be
(\delta \calE_\zeta^\star)^2=\sigma_{\rm prior}^2-V_2^2/V_1.
\ee
We can see that the variance decrease $V_2^2/V_1$ is large when the variance of the outcome is small and the cross correlation is large. 
With the overall precision in hand, to characterize the precision of the second measurement $\delta q^{\rm GKP}_\eta$, we calculate the Fisher information increase 
$
1/(\delta q^{\rm GKP}_\eta)^2=1/(\delta \calE_\zeta^\star)^2-1/\sigma_{\rm prior}^2,
$
thus the effective measurement rms precision
\be
\delta q^{\rm GKP}_\eta=\sigma_{\rm prior}\sqrt{\frac{\sigma_{\rm prior}^2}{V_2^2/V_1}-1}.
\ee 
To evaluate the precision, we will consider $\sigma_{\rm prior}^2=k_{\rm prior}/4M^2n_S$ with various $k_{\rm prior}$. As shown in Fig.~\ref{GKP_M}, we do see a Heisenberg-scaling precision, regardless of the different values of $k_{\rm prior}$. Especially, when $k_{\rm prior}\sim 1$, we see a good agreement with the performance without the modulo $\sqrt{2\pi}$ complication, $\delta q^{\rm GKP}_\eta\sim 1/(2M \sqrt{n_S})$.

Finally we address the loss issue and possible generalizations.
When there is loss, one can utilize the CV error correction schemes to mitigate the loss, similar to the scheme analyzed in Section~\ref{Sec_improvement_real}. 


Due to the constraints from the modulo operations, we are restricted to the equal-weight scenario in the previous analysis. The generalization to the unequal-weight scenario is possible, at the cost of introducing excess noises. Suppose we keep the beamsplitters balanced and the post-processing in the estimator the same as in Fig.~\ref{DS_GKP_no_loss}; we can still tune the weights $\{w_m, 1\le m \le M\}$ by concatenating each displacement unitary with a suitable loss or amplification channel. Consider the rescaled weights, $\{k_m\equiv M w_m, 1\le m \le M\}$, with the normalization $\sum_{m=1}^M k_m=M$. When $k_m>1$, one can apply a loss $\calL_{1/k_m^2}$ and a gain $\calA_{k_m^2}$ such that the combined channel 
\be
\calA_{k_m^2}
\circ \hat{U}(\alpha_m)
\circ \calL_{1/k_m^2}=\Phi_{\sqrt{k_m^2-1}}\circ \hat{U}(k_m\alpha_m),
\ee
where the notation $\hat{U}(\cdot)$ is now a unitary displacement channel. When $k_m<1$, one can apply a gain $\calA_{1/k_m^2}$ first and then loss $\calL_{k_m^2}$, such that the combined channel
\be
\calL_{k_m^2}
\circ \hat{U}(\alpha_m)
\circ \calA_{1/k_m^2} =\Phi_{\sqrt{1-k_m^2}}\circ \hat{U}(k_m\alpha_m).
\ee
With the above equivalence relation, afterwards an equal-weight addition of the measurement results will give the correct weighted average of displacements, up to some overall noise with variance $\sum_{m=1}^M  |k_m^2-1|/M$.

\section{Discussion}
In principle, continuous-variable error correction codes such as 
GKP-stabilizer codes may be used to enhance the reliability of any protocol that makes use of CV quantum information.
In this paper, we focus on the enhancement of distributed sensing tasks that can be achieved with CV error correction, providing a detailed evaluation of the effectiveness of GKP-stabilizer codes used for this purpose. When used for distributed sensing of CV displacements, the GKP-stabilizer code with six iterations (level $n=7$ code) reinstates Heisenberg-scaling of precision up to about $10^2$ nodes for transmissivity $\eta\geq 0.95$, while the Heisenberg scaling is destroyed entirely when no error correction is used. Since the GKP grid state enables simultaneous precise measurements of small displacements on both quadratures, we also use it to extend the distributed sensing protocol from single-quadrature displacements to displacements of both quadratures. When good prior information is available, simultaneous Heisenberg scaling of rms estimation errors on both quadratures can be achieved for the equal-weight case. 

Three future directions are worth pointing out. First, there is room for improvement on the conversion from a pure loss channel to an AWGN channel in section~\ref{Sec_EC}. During the amplification, part of the information about the input state is stored in the environment mode, and it might improve performance if the environment mode is used in the decoding process as well. It will be worthwhile to investigate whether using CV error correction in quantum repeaters will improve their performance against loss. Finally, our methods can be applied to sensing other parameters, such as a weighted average of phases, where we expect similar enhancements of performance.

\begin{acknowledgements}
We acknowledge Kyungjoo Noh and Sisi Zhou for discussions. We acknowledge support from the University of Arizona, ARO (W911NF-19-1-0418), ONR (N00014-19-1-2189), ARO-LPS (W911NF-18-1-0103), NSF (PHY-1733907), ARL-CDQI (W911NF-15-2-0067), ARO (W911NF-18-1-0020, W911NF-18-1-0212), ARO MURI (W911NF-16-1-0349), AFOSR MURI (FA9550-15-1-0015, FA9550-19-1-0399), DOE (DE-SC0019406), NSF (EFMA-1640959, OMA-1936118), and the Packard Foundation (2013-39273). The Institute for Quantum Information and Matter is an NSF Physics Frontiers Center. Q.Z. acknowledges the hospitality of the Yale Quantum Institute during the completion of the paper.
\end{acknowledgements}


%

\end{document}